\documentstyle{article}

\begin{document}

\title{\flushright{\small KL-TH/ 02-07} \bigskip \bigskip \bigskip \bigskip
\bigskip \bigskip \\
\center{The Jordanian Bicovariant Differential Calculus}\thanks{%
To appear in the proceedings  of the XXIV International Colloquium on Group Theoretical Methods
in Physics, Paris 15-20 July 2002.}}
\author{L. MESREF\thanks{%
Email: lmesref@physik.uni-kl.de}}
\date{Department of physics, Theoretical Physics\\
University of Kaiserslautern, Postfach 3049\\
67653 Kaiserslautern, Germany}
\maketitle

\begin{abstract}
We show that the Woronowicz prescription using a bimodule constructed out of
a tensorial product of a bimodule and its conjugate and a bi-coinvariant
singlet leads to a trivial differential calculus.
\end{abstract}

\section{Introduction}

It is by now well known that our naive conception of the space-time as a
collection of points equipped with suitable topological and metric
structures at the energies much below the Planck scale should be modified.
One possible approach to the description of physical phenomena at small
distance is based on non-commutative geometry of the space-time \cite{connes}%
. In the quantum groups picture \cite{drinfeld} the symmetry is described by
noncommutative non-cocommutative $\ast $ Hopf algebra. The connection with \
noncommutative differential geometry has been made by Woronowicz \cite
{woronowicz} \ who introduced the theory of bicovariant diferential
calculus. This turns out to be the appropriate way to describe quantum gauge
theories. In this letter we show that this method leads to a trivial
calculus in the case of the Jordanian group $U_{h}\left( 2\right) $.

\section{The Jordanian Quantum Group $U_{h}\left( 2\right) $}

We recall that there are only two quantum group structures which admit a
central determinant on space of $2\times 2$ matrices: $GL_{q}\left( 2\right) 
$ \cite{takhtajan} and $GL_{h}\left( 2\right) $ \cite{demidov} (the
deformation of $M\left( 2\right) $ was considered and named 
\'{}%
`Jordanian'' by Manin \cite{manin}). The continuous parameter $h$ was
introduced by Zakrzewski \cite{zakrzewski}.

Let $\mathcal{A}$ be the associative unital $C$-algebra generated by the
linear transformations $M^{n}\,_{m}$ $\left( n,m=1,2\right) $

$.$

\begin{equation}
M^{n}\,_{m}=\left( 
\begin{array}{cc}
a & b \\ 
c & d
\end{array}
\right) ,
\end{equation}

\bigskip

the elements $a$, $b$, $c$, $d$ satisfying the relations

\bigskip

\begin{eqnarray}
\left[ a,c\right] &=&hc^{2},\qquad \ \left[ b,a\right] =h\left(
a^{2}-D_{h}\right) ,    \nonumber \\
\left[ d,c\right] &=&hc^{2},\qquad \ \left[ d,b\right] =h\left(
D_{h}-d^{2}\right) ,     \nonumber  \\
\left[ a,d\right] &=&h\left( d-a\right) c,\qquad \ \left[ b,c\right]
=h\left( ac+cd\right) ,
\end{eqnarray}

\bigskip

where $D_{h}=ad-cb-hcd.=ad-bc+hac$ is the Jordanian central determinant. The
classical case is obtained by setting h equal to zero. The relations (2) are
obtained by applying either the method of Faddeev et al. \cite{reshetikhin}
namely by solving the monodromy equation $R\,M_{1}M_{2}=M_{2}M_{1}R$ where $%
M_{1}=M\otimes I,\,M_{2}=I\otimes M$ \ and $R$ is given in Eq. (6) or the
method of Manin \cite{manin} using $M$ as transformation matrix of the
appropriate quantum planes.

The $U_{h}\left( 2\right) $ is obtained by requiring that the unitary
condition hold for this $2\times 2$ matrix:

\bigskip

\begin{equation}
M_{m}^{n\ \dagger }=M_{m}^{n\ -1}.
\end{equation}

\bigskip

The $2\times 2$ matrix belonging to $U_{h}\left( 2\right) $ preserves the \
nondegenerate bilinear form $B_{nm}$ \cite{dubois}

\bigskip

\begin{equation}
B_{nm}M_{k}^{n}M_{l}^{m}=D_{h}B_{kl},\quad
B^{nm}M_{n}^{k}M_{m}^{l}=D_{h}B^{kl},\quad B_{kn}B^{nl}=\delta _{k}^{l},
\end{equation}

\bigskip

\begin{equation}
B_{nm}=\left(
\begin{array}{cc}
0 & -1 \\
1 & h
\end{array}
\right) ,\qquad B^{nm}=\left( 
\begin{array}{cc}
h & 1 \\ 
-1 & 0
\end{array}
\right) ,\quad B^{nm}B_{nm}=-2 .
\end{equation}

\bigskip

\section{$U_{h}\left( 2\right) $ Woronowicz Bicovariant Differential Calculus}

Zakrzewski \cite{zakrzewski} has applied the general construction of the
Leningrad School \cite{faddeev} to the following $R$ matrix which controls
the noncommutativity of the elements $M_{\,\,\,m}^{n}$

\bigskip

\begin{equation}
R=\left( 
\begin{array}{cccc}
1 & -h & h & h^{2} \\ 
0 & 0 & 1 & -h \\
0 & 1 & 0 & h \\ 
0 & 0 & 0 & 1
\end{array}
\right) .
\end{equation}

\bigskip

The $R$ matrix becomes the permutation operator $R_{\qquad kl}^{nm}=\delta
_{l}^{n}\delta _{k}^{m}$ in the classical limit $h=0.$

The braiding $R$ matrix satisfy the Yang-Baxter equation

\bigskip

\begin{equation}
R_{\,\,\,pq}^{ij}R_{\,\,\,\,\,\,lr}^{pk}R_{\,\,\,\,\,\,mn}^{qr}=R_{\,\,%
\,pq}^{jk}R_{\,\,\,\,\,\,rm}^{ip}R_{\,\,\,\,\,\,lm}^{rq} .
\end{equation}

\bigskip

The noncommutativity \ of the elements $M_{\,m}^{n}$ is expressed as

\bigskip

\begin{equation}
R_{\,\,\,\,\,\,\,nm}^{pq}M_{\,\,\,k}^{n}M_{\,\,\,l}^{m}=M_{\,\,\,n}^{p}M_{\,%
\,\,m}^{q}R_{\,\,\,\,\,\,\,\,kl}^{nm}. 
\end{equation}

\bigskip

The algebra $Fun\left( U_{h}\left( 2\right) \right) $ is a Hopf algebra with
comultiplication $\Delta $, counit $\epsilon $ and antipode $S$ which are
given by:

\bigskip - \ comultiplication (also called coproduct )

\bigskip

\begin{equation}
\Delta \left( M_{\,\,m}^{n}\right) =M_{\,\,k}^{n}\otimes M_{\,\,m}^{k} .
\end{equation}

\bigskip

This coproduct $\Delta $ on $Fun\left( U_{h}\left( 2\right) \right) $ is
directly related, for $h=0$ (the non deformed case), to the pullback induced
by left multiplication of the group on itself.

\bigskip - \ co-unit $\epsilon $

\begin{equation}
\varepsilon \left( M_{\,\,m}^{n}\right) =\delta _{m}^{n},
\end{equation}

- \ antipode $S$\ (coinverse)\ \ \ \ \ \ \ \

\bigskip

\begin{equation}
\ \ S\left( M_{\,\,k}^{n}\right) M_{\,\,m}^{k}=M_{\,\,k}^{n}S\left(
M_{\,\,m}^{k}\right) =\delta _{\,m}^{n},
\end{equation}

\bigskip

\ \ \ \ \ \ \ \ \ \ \ \ \ \ \ \ \ \ \ \ \ \ \ \ \ 
\begin{equation}
S\left( M_{\,m}^{n}\right) =\frac{1}{D_{h}}B^{nk}M_{\,k}^{l}B_{lm} .
\end{equation}

\bigskip

With the nondegenerate bilinear form $B$, the $R$ matrix has the form

\bigskip

\begin{eqnarray}
R_{\qquad kl}^{+nm} &=&R_{\qquad kl}^{nm}=\delta _{\quad k}^{n}\delta
_{\quad l}^{m}+B^{nm}B_{kl},      \nonumber \\
R_{\qquad kl}^{-nm} &=&R_{\qquad \ \ kl}^{-1nm}=R_{\qquad kl}^{nm}.
\end{eqnarray}

\bigskip

The $R$ matrix satisfies the Hecke relations $R^{\pm 2}=1$ and the relation

\bigskip

\begin{equation}
B_{nm}R_{\qquad kc}^{an}R_{\qquad lb}^{cm}=\delta _{b}^{a}B_{kl}. 
\end{equation}

\bigskip

Now, we are going to consider the bicovariant bimodule $\Gamma $ over $%
U_{h}\left( 2\right) $. Let $\theta ^{a}$ be a right invariant basis of $%
\Gamma _{inv},$ the linear subspace of all right -invariant elements of $%
\Gamma $ i.e. $\Delta _{R}\left( \theta ^{a}\right) =\theta ^{a}\otimes I.$
In the $h=0$ the right coaction $\Delta _{R}$ coincides with the pullback
for 1-forms. The left action is defined as

\bigskip 

\begin{equation}
\Delta _{L}\left( \theta ^{a}\right) =M_{\,\,\,b}^{a}\otimes \theta ^{b} .
\end{equation}

In the Jordanian quantum case we have $\theta ^{a}M_{\,\,\,m}^{n}\neq
M_{\,\,\,m}^{n}\theta ^{a}$ in general, the bimodule structure of $\Gamma $
being non-trivial for $h\neq 0.$ There exist linear functionals $\
f_{\,\,\,b}^{a}:Fun\left( U_{h}\left( 2\right) \right) \rightarrow \mathcal{C%
}$ for these left invariant basis such that

\bigskip

\begin{equation}
\theta ^{a}M_{\,\,\,m}^{n}=\left( M_{\,\,\,m}^{n}\ast \
f_{\,\,\,b}^{a}\right) \theta ^{b}=\left( f_{\,\,\,b}^{a}\otimes id\right)
\Delta \left( M_{\,\,\,m}^{n}\right) \theta ^{b}=f_{\,\,\,b}^{a}\left(
M_{\,\,\,k}^{n}\right) M_{\,\,\,m}^{k}\theta ^{b} .
\end{equation}

\bigskip

Once we have the functionals $f_{\,\,\,b}^{a}$, we know how to commute
elements of $\mathcal{A}$ through elements of $\Gamma $. These functionals
satisfy the consistent conditions:

\bigskip

\begin{eqnarray}
f_{\,\,\,b}^{a}\left( M_{\,\,\,m}^{n}M_{\,\,\,l}^{k}\right)
&=&f_{\,\,\,c}^{a}\left( M_{\,\,\,m}^{n}\right) \,f_{\,\,\,b}^{c}\left(
M_{\,\,\,l}^{k}\right)           \nonumber  \\
f_{\,\,\,b}^{a}\left( I\right) &=&\delta _{b}^{a}   \nonumber \\
\left( f_{\,\,\,c}^{a}\circ S\right) \,f_{\,\,\,b}^{c} &=&\delta
_{b}^{a}\,\epsilon ;\quad f_{\,\,\,c}^{a}\,\left( f_{\,\,\,b}^{c}\circ
S\right) =\delta _{b}^{a}\,\epsilon .
\end{eqnarray}

\bigskip

Using these conditions, we find from Eq. (4) and Eq. (14) $\ f_{\quad
b}^{a}\left( M_{\quad k}^{n}\right) =\left( D_{h}\right) ^{\frac{1}{2}%
}R_{\qquad kb}^{an}.$

We can also define the conjugate basis $\theta ^{\ast a}=\left( \theta
^{a}\right) ^{\ast }\equiv \overline{\theta }_{a}.$

The left coaction acts on these basis as

\bigskip

\begin{equation}
\Delta _{L}\left( \overline{\theta }_{a}\right) =S\left(
M_{\,\,\,\,a}^{b}\right) \otimes \overline{\theta }_{b}.
\end{equation}

\bigskip

This equation is easily obtained from Eq. (15) by the antilinear $\ast $\
involution using the relations $\left( \Delta _{R}\left( \theta ^{a}\right)
\right) ^{\ast }=\Delta _{R}\left( \theta ^{a}\right) ^{\ast },\left(
M_{\,\,\,m}^{n}\right) ^{\ast }=M_{\,\,\,\,\,\,m}^{\ast n}\equiv
M_{\,\,\,\,n}^{\dagger m}=S\left( M_{\,\,\,n}^{m}\right) .$ Then the linear
functionals $\overline{f}_{\,\,\,b}^{a}$ are given by

\bigskip

\begin{equation}
\overline{\theta }_{b}M_{\,\,m}^{n}=\left( \ M_{\,\,\,m}^{n}\ast \overline{f}%
_{\,\,\,\,b}^{a}\right) \overline{\theta }_{a},
\end{equation}

\bigskip

where the functionals for the conjugated basis $\overline{\theta }_{a}$ is
given by:

\bigskip

\begin{equation}
\overline{f}_{\,\,\,b}^{a}\left( S\left( M_{\,\,m}^{n}\right) \right)
=\left( D_{h}\right) ^{\frac{-1}{2}}R_{\,\,\,\,\,\,\,\,\,\,\,\,mb}^{-an}.
\end{equation}

\bigskip

The representation with the upper index of $\overline{\theta }^{a}$ is
defined by using the nondegenerate bilinear form B:

\bigskip
\begin{equation}
\overline{\theta }^{b}=\overline{\theta }_{a}B^{ab}.
\end{equation}

\bigskip This gives 
\begin{equation}
\Delta _{L}\left( \overline{\theta }^{a}\right) =M_{\,\,\,\,b}^{\,a}\otimes 
\overline{\theta }^{b},
\end{equation}

\bigskip

which defines the new functionals $\widetilde{f}_{\,\,\,b}^{a}$
corresponding to the basis $\overline{\theta }^{a}$

\bigskip
\begin{equation}
\widetilde{f}_{\,\,\,b}^{a}=B_{bc}\overline{f}_{\,\,\,\,\,d}^{c}B^{da}.
\end{equation}

\bigskip

We can easily find the transformation of the adjoint representation for the
Jordanian quantum group which acts on the generators $M_{\,\,m}^{n}$ as the
left coaction $Ad_{L}$:

\bigskip

\begin{equation}
Ad_{L}\left( M_{\,\,\,m}^{n}\right) =M_{\,\,\,l}^{n}\,S\left(
M_{\,\,\,\,m}^{k}\right) \otimes M_{\,\,\,k}^{l}.
\end{equation}

\bigskip

As usual, in order to define the bicovariant differential calculus with the $%
\ast -$structure we have to require that the $\ast -$operation is a bimodule
antiautomorphism
$\left( \Gamma _{Ad}\right) ^{\ast }=\Gamma _{Ad}$. We find
the right invariant bases containing the adjoint representation. They are
obtained by taking the tensor product $\theta ^{a}\overline{\theta }%
_{b}\equiv \theta _{\,\,b}^{a}$ of two fundamental modules. The bimodule
generated by these bases is closed under the $\ast -$operation. Using the
fact that $\Delta _{L}\left( \theta ^{a}\overline{\theta }_{b}\right)
=\Delta _{L}\left( \theta ^{a}\right) \Delta _{L}\left( \overline{\theta }%
_{b}\right) $ we find the left coaction on the basis $\theta _{\,\,b}^{a}$

\bigskip

\begin{equation}
\Delta _{L}\left( \theta _{\,\,\,b}^{a}\right) =S\left(
M_{\,\,\,\,c}^{a}\right) M_{\,\,\,\,b}^{d}\otimes \theta _{\,\,\,d}^{c}.
\end{equation}

\bigskip

In this basis the left coaction is given by

\bigskip

\begin{equation}
\Delta _{L}\left( \theta ^{ab}\right) =\frac{1}{D_{h}}M_{\,\,\,\,c}^{a}M_{\,%
\,\,\,d}^{b}\otimes \theta ^{cd}.
\end{equation}

\bigskip

We can deduce the relation between the left and the right multiplication for
this basis

\bigskip

\begin{equation}
\theta ^{ab}M_{\,\,\,m}^{n}=\left( M_{\,\,\,m}^{n}\ \ast
f_{Ad\,\,\,\,\,\,\,\,cd}^{\,\,\,\,\,\,\,\,\,ab}\right) \theta
^{cd}=f_{Ad\,\,\,\,\,\,\,\,cd}^{\,\,\,\,\,\,\,\,\,ab}\left(
M_{\,\,\,k}^{n}\right) M_{\,\,\,m}^{k}\theta ^{cd},
\end{equation}

\bigskip

where

\begin{equation}
f_{Ad\,\,\,\,\,\,\,\,cd}^{\,\,\,\,\,\,\,\,\,ab}=\widetilde{f}%
_{\,\,\,d}^{b}\ast f_{\,\,\,c}^{a}. 
\end{equation}

\bigskip

The exterior derivative d is defined as

\bigskip

\begin{eqnarray}
dM_{\,\,\,m}^{n} &=&\frac{1}{\mathcal{N}}\left[ X,M_{\,\,\,m}^{n}\right]
_{-}=\,\theta ^{ab}\left( M_{\,\,\,m}^{n}\ast \chi _{ab}\right)   \nonumber \\
&=&\chi _{ab}\left( M_{\,\,\,k}^{n}\right) \theta ^{ab}M_{\,\,\,m}^{k},
\end{eqnarray}

\bigskip

where $X=B_{ab}\theta ^{ab}=-\theta ^{12}+\theta ^{21}+h\theta ^{22}$ is the
singlet representation of $\theta ^{ab}$ and is both left and right
co-invariant, $\mathcal{N}\in \mathcal{C}$ is the normalization constant
which we take purely imaginary $N^{\ast }=-N$ and $\chi _{ab}$ are the
quantum analogue of right- invariant vector fields.

Using (20),(23), (28)

\bigskip

\begin{eqnarray}
dM_{\,\,\,m}^{n} &=&\frac{1}{\mathcal{N}}\left( B_{ab}\delta
_{m}^{k}M_{\,\,\,k}^{n}\theta
^{ab}-B_{cd}f_{Ad\,\,\,\,\,\,\,\,ab}^{\,\,\,\,\,\,\,\,\,cd}\left( S\left(
M_{\,\,\,m}^{k}\right) \right) M_{\,\,\,k}^{n}\theta ^{ab}\right) \nonumber  \\
&=&\frac{1}{\mathcal{N}}\left( B_{ab}\delta
_{m}^{k}-B_{cd}R_{\,\,\,\,\,\,ma}^{ct}R_{\,\,\,\,\,\,\,\,tb}^{dk}\right)
M_{\,\,\,k}^{n}\theta ^{ab}=0.
\end{eqnarray}

\bigskip

From (29) and (30) we deduce

\bigskip

\begin{equation}
\chi _{ab}\left( M_{\,\,\,m}^{k}\right) =0. 
\end{equation}

\bigskip

We \ see that it is a trivial calculus $dM_{\,\,\,m}^{n}=0$. To obtain a
nontrivial calculi we have  followed, in a recent paper \cite{mes}, the
Karimipour \cite{karimipour} method for our $4D$ calculus and constructed
a Jordanian trace. This trace has permitted us to define an invariant $U_{h} \left( 2\right) $ Yang-Mills Lagrangian. The Jordanian BRST and anti-BRST transformations \cite{mesre} can also be carried out and will be reported in a future work.

\bigskip 

\textbf{Acknowledgments}

I would like to thank the D.A.A.D. for its financial support and Prof. Dr. W. Ruhl for reading the manuscript and encouragement.

\bigskip

\end{document}